\documentclass[%
 footinbib,
 reprint,
superscriptaddress,
longbibliography,
 amsmath,amssymb,
 aps,
 floatfix,
 color,
 nobalancelastpage,
]{revtex4-1}

\usepackage{braket}
\usepackage{tikz}
\usepackage{dsfont}

\usepackage[version=4]{mhchem}

 \usepackage{pgfplots}
  \pgfplotsset{compat=newest}

\usetikzlibrary{external}
\usetikzlibrary{calc,  shapes, backgrounds, arrows, chains, matrix, positioning, scopes,
                decorations, decorations.pathmorphing,patterns, fit, pgfplots.groupplots, plotmarks}
\tikzset{>=latex}
\usetikzlibrary{external}
\usepackage{graphicx}% Include figure files
\usepackage{dcolumn}% Align table columns on decimal point
\usepackage{bm}% bold math

\usepackage{hyperref}

\begin{document}

\preprint{APS/123-QED}

\title{Fork Tensor Product States - Efficient Multi-Orbital Real Time DMFT Solver}

\author{Daniel Bauernfeind}
\email[]{daniel.bauernfeind@tugraz.at}
\affiliation{%
  Institute of Theoretical and Computational Physics\\
  Graz University of Technology, 8010 Graz, Austria
}%

\author{Manuel Zingl}
\affiliation{%
  Institute of Theoretical and Computational Physics\\
  Graz University of Technology, 8010 Graz, Austria
}%

\author{Robert Triebl}
\affiliation{%
  Institute of Theoretical and Computational Physics\\
  Graz University of Technology, 8010 Graz, Austria
}%

\author{Markus Aichhorn}
\affiliation{%
  Institute of Theoretical and Computational Physics\\
  Graz University of Technology, 8010 Graz, Austria
}%

\author{Hans Gerd Evertz}
\affiliation{%
  Institute of Theoretical and Computational Physics\\
  Graz University of Technology, 8010 Graz, Austria
}%
\affiliation{%
  Kavli Institute for Theoretical Physics \\ 
  University of California, Santa Barbara, CA 93106, USA
}%

\date{\today}

\begin{abstract}
We present a tensor network especially suited for multi-orbital
Anderson impurity models and as an impurity solver for
multi-orbital dynamical mean-field theory (DMFT). The solver works directly on the
real-frequency axis and yields high spectral
resolution at all frequencies. We use a large number $\left( \mathcal{O}(100) \right )$ of bath sites, 
and therefore achieve an accurate representation of the bath. The solver can treat
full rotationally-invariant interactions with reasonable numerical effort. We
show the efficiency and accuracy of the method by a benchmark for the
three-orbital testbed material \ce{SrVO3}. There we observe multiplet structures in the high-energy
spectrum which are almost impossible to resolve by other multi-orbital methods. The
resulting structure of the Hubbard bands can be described as a broadened atomic 
spectrum with rescaled interaction parameters. Additional features emerge when $U$ is increased.
Finally we show that our solver can be applied even to models with
five orbitals.
This impurity solver offers a new route to the calculation
of precise real-frequency spectral functions of correlated materials.
%\begin{description}
%\end{description}
\end{abstract}

\maketitle

\section{\label{sec:intro}Introduction}

Strongly correlated systems are among the most fascinating objects
solid-state physics has to offer. The interactions between constituents of such
systems lead to emergent phenomena that cannot be deduced from the
properties of non-interacting particles~\cite{Anderson_MoreIsDiff}.

One of the most widely used methods to describe strongly-correlated
electrons is the dynamical mean-field theory
(DMFT)~\cite{GeorgesDMFT,MetznerVollhardt_Dinf}. DMFT treats local
electronic correlations by a self-consistent mapping of the lattice
problem onto an effective 
Anderson impurity model (AIM). Calculating the single particle spectral function of this impurity model in an
accurate and efficient way is at the heart of every DMFT
calculation. To this end, many numerical methods have been developed or
adapted. These are based for instance on continuous-time quantum Monte Carlo
(CTQMC)~\cite{GullCTQMC,WernerFirstCTQMC}, exact diagonalization
(ED)~\cite{Caffarel_EDIMPSOLVER, CaponeEDDMFT,LichtensteinED}, the numerical
renormalization group (NRG)~\cite{WilsonNRG, BullaNRG}, configuration interaction (CI)
based solver~\cite{HaverkortED,ZgidCI}, and also 
the density-matrix renormalization group (DMRG) with matrix-product states
(MPS)~\cite{WhiteDMRG,SchollwoeckDMRG_MPS}. 

Every algorithm has strengths and weaknesses: CTQMC is exact apart from statistical errors
on the imaginary axis and can deal with multiple orbitals, but it is in
some cases plagued by the fermionic sign problem. Additionally, an ill-posed
analytic continuation is necessary to obtain real-frequency
spectra, which therefore become broadened, especially at high energies.
 ED directly provides spectra on the real axis, but it is severely
limited in the size of the Hilbert space, i.e.\ in the number of bath
sites. Quite recently, NRG was shown to be a viable three-band solver
by exploiting non-abelian quantum number
conservation~\cite{Stadler_NRG3Band, Jernej_NRG3Band, VanDelft_InterleavedNRG}. NRG 
works on the real axis and captures the low-energy physics well, but it
has by construction a poor resolution at higher energies. 
Another interesting route that has been proposed recently are solvers
that tackle the problem of exponential growth of the Hilbert space
using ideas from quantum chemistry, i.e.~the configuration
interaction~\cite{HaverkortED,ZgidCI}. They allow to go beyond the
small bath sizes of ED, keeping all the advantages such as absence of
fermionic sign problems. However, in multi-orbital applications (see
Appendix of Ref.~\cite{ZgidCI}), the spectral resolution has so far been 
restricted by the restricted number of bath sites ($\mathcal{O}(20)$).

MPS based techniques like DMRG, finally, do not suffer from a sign problem and can be used on
the real- as well as on the imaginary-frequency axis. The price to pay for the absence of the 
sign problem is an, in general, very large growth of bond dimension with the number of orbitals.

Dynamical properties and spectral functions can be calculated within
DMRG and have been used for impurity solvers, e.g.\ with the Lanczos-like continued-fraction
expansion~\cite{HallbergSpecFunc,Garcia_DMFTwithMPS}.
Other solvers using the more stable correction vector~\cite{WhiteDynCorrFunc} 
and dynamical DMRG (DDMRG)~\cite{JeckelmannDDMRG} methods were developed~\cite{Nishimoto_IMPSOLVER_MPS,PetersSpecFunc,KarskiMOTTDMFT,KarskiMOTTDMFT2}.
Both algorithms produce very accurate spectral functions, but have the disadvantage that a separate calculation for each
frequency has to be performed. 
The Chebyshev expansion~\cite{WeisseKPM} with MPS~\cite{HolznerCheby}, 
supplemented by linear prediction~\cite{WhiteLinPred}, was used for
impurity solvers in the single band case~\cite{GanahlCheby} and for two bands~\cite{WolfCheby}.
Recently, some of us introduced a method based on real-time evolution~\cite{VidalTEBD1,VidalTEBD2,WhiteFeiguinTDMRG,SchollwoeckTDMRG}
and achieved a self consistent DMFT solution for a two-band model~\cite{Ganahl2BHubb}.
In such calculations, the physical orbitals for each spin direction are usually combined to
one large site in the MPS. Three or more orbitals  
have not been feasible with this approach, because of a large increase in
computational cost with the number of orbitals. 
Another MPS-based solver, which works on the imaginary axis, was recently introduced~\cite{Wolf_ImagTimeSolver}
and it was applied as a solver for three bands in two-site cluster DMFT.
It was supplemented by a single real-time evolution to compute the spectral function,
avoiding the analytic continuation. However, this method is
restricted by the number of bath sites which can be employed. In the calculation mentioned,
only three bath sites per orbital were used, limiting the energy resolution for real-frequency 
spectral functions.

In the present paper, we introduce a novel impurity solver which works directly
on the real-frequency axis.
To this end, we use a tensor network that captures the
geometry of the interactions in the Anderson model better than a
standard MPS. Our approach is to some extent inspired by the work of Ref.~\cite{HolznerForkMPS},
which used a similar network for a two orbital NRG ground state calculation.
We are not restricted to a small number of bath sites.
This is imperative for exploiting the spectral resolution achievable 
with real-time calculations. We emphasize that (i) our method is
by construction free of any fermionic sign problem, (ii) one can
fully converge the DMFT self-consistency loop on the real-frequency 
axis and (iii) we can achieve an almost exact representation of the bath spectral
function. We apply this method to multi-orbital DMFT for the testbed
material \ce{SrVO3} and show that one can resolve a multiplet
structure in the Hubbard bands, keeping at the same time a good
description of the low-energy quasi-particle excitations.

The paper is structured as follows. First we show how impurity solvers
with tensor networks work in general and introduce our new tensor
network approach which we call fork tensor-product states (FTPS)
(Sec. \ref{sec:SolverTensNet}). Next we explain in detail how our
solver is used in the context of multi-orbital DMFT
(Sec. \ref{sec:DMFT}). In Sec. \ref{sec:Results}, we apply our
approach to \ce{SrVO3} and discuss the multiplet structure that the
FTPS solver allows to resolve. In order to check the accuracy of the
method, we also compare the FTPS results to CTQMC for
\ce{SrVO3}. Finally, we show the efficiency of the FTPS solver by
applying it to a five-orbital model.

\section{\label{sec:SolverTensNet}Tensor Network Impurity Solvers}
The Anderson impurity model (AIM) describes an impurity (with Hamiltonian
$H_{\text{loc} }$) coupled to a bath of non-interacting fermions hybridized
with it. A typical AIM Hamiltonian is given by:

 \begin{align}\label{eq:H_DMFT3B}
	H &=   H_{\text{loc}} + H_{\text{bath}}  \\
	H_{\text{loc}} &= \epsilon_0 \sum_{m \sigma} n_{m 0 \sigma} + H_{ \text{DD} } + H_{\text{SF-PH} }\nonumber \\
	H_{\text{DD}} &=  U \sum_m n_{m0\uparrow} n_{m0\downarrow}
        \nonumber \\
        &+(U-2J) \sum_{ m'>m, \sigma } n_{m0\sigma} n_{m' 0\bar{\sigma}} \nonumber\\
	  &+(U-3J) \sum_{ m'>m, \sigma } n_{m0\sigma} n_{m' 0 \sigma} \nonumber \\	  
	H_{ \text{SF-PH} } &= J\sum_{m'>m} \left ( c_{m 0 \uparrow
          }^{\dag} c_{m 0 \downarrow }    c_{m' 0 \uparrow } c_{m' 0
            \downarrow }^{\dag} +  \text{h.c.} \right ) \nonumber\\ 
	 	  &-J\sum_{m'>m} \left( c_{m 0 \uparrow }^{\dag} c_{m 0 \downarrow }^{\dag} c_{m' 0 \uparrow } c_{m' 0 \downarrow } + \text{h.c.}  \right )  \nonumber\\
 	H_{ \text{bath} } &= \sum_{m l \sigma} \epsilon_l n_{ml\sigma} + V_l \left ( c_{m0\sigma}^{\dag} c_{ml\sigma} + \text{h.c.} \right ),\nonumber
\end{align}

where $c_{m l \sigma}^{\dag}$ ($c_{m l \sigma}$) creates (annihilates) an
electron in band $m$ ($m \in \{ 1,2,3 \}$ for a three-orbital model) with spin $\sigma$ at the
$l$-th site of the system (the impurity has index $l=0$, the bath
degrees of freedom have $l\geq 1$), and $n_{m l \sigma}$
are the corresponding particle number operators. $H_{\text{DD} }$
describes density-density (DD) interactions between all orbitals and
$H_{ \text{SF-PH} }$ are the spin-flip and pair-hopping terms. This
three-orbital Hamiltonian is not only important in the context of
real-material calculations. It has also been studied extensively on
the model level, most importantly because it hosts unconventional
correlation phenomena. For a selection of recent work, see for
instance
Refs.~\cite{werner_spinfreezing,Stadler_NRG3Band,demedici2011,GeorgesHundJ,
PhysRevLett.118.086401}.

An impurity solver calculates the retarded impurity Greens function $G(t)$
\begin{equation} \label{eq:GF}
  G(t) = -i \theta(t) \bra{\psi_0} [ c^{\dag}(t), c(0) ] \ket{\psi_0}
\end{equation}
of the interacting problem~\eqref{eq:H_DMFT3B}, either in real or
imaginary time $t$. In the present paper,
we introduce a new tensor network similar to an MPS, which can be used as
a real-time impurity solver for three orbitals. We first
introduce MPS before moving on to what we call
fork tensor-product states (FTPS) in Sec. \ref{sec:FORKMPS}.

\subsection{\label{sec:MPS}Matrix Product States (MPS) and DMRG}
MPS are a powerful tool to efficiently encode quantum mechanical
states. Consider a state $\ket{\psi}$ of a system consisting
of $N$ sites: 
\begin{equation}\label{eq:state}
\ket{\psi} = \sum_{s_1,s_2, \cdots, s_N} c_{s_1, \cdots, s_N} \ket{s_1, s_2, \cdots, s_N}\text{.}
\end{equation}
Each site $i$ has a local Hilbert space  of dimension $d_i$ spanned by
the states $\ket{s_i}$. 
Through repeated use of singular-value decompositions (SVDs), it is
possible to factorize every coefficient $c_{s_1, \cdots, s_N}$
into a product of matrices~\cite{SchollwoeckDMRG_MPS}, i.e.\ into an
MPS,  
\begin{equation}\label{eq:defMPS}
\ket{\psi} = \sum_{s_1,s_2, \cdots, s_N} A_1^{s_1} \cdot  A_2^{s_2} \cdots  A_N^{s_N}  \ket{s_1, s_2, \cdots, s_N}.
\end{equation}
Each $A_i^{s_i}$ is a rank-3 tensor, except the first and last ones
($A_1^{s_1}$, $A_N^{s_N}$), which are of rank two. The index $s_i$ is
called physical index, and the \emph{matrix} indices, which are summed
over, are the so called bond indices. A general state of the full Hilbert
space is unfeasible to store, but it can be shown that ground states
are well described by an MPS with limited bond dimension $m$
(dimension of the bond index)~\cite{VerstraeteGSrepFaithf}. 

\begin{figure}[ht]
   \centering
    %\scalebox{1}{ \input{tikzfigs/mps} }
    \includegraphics{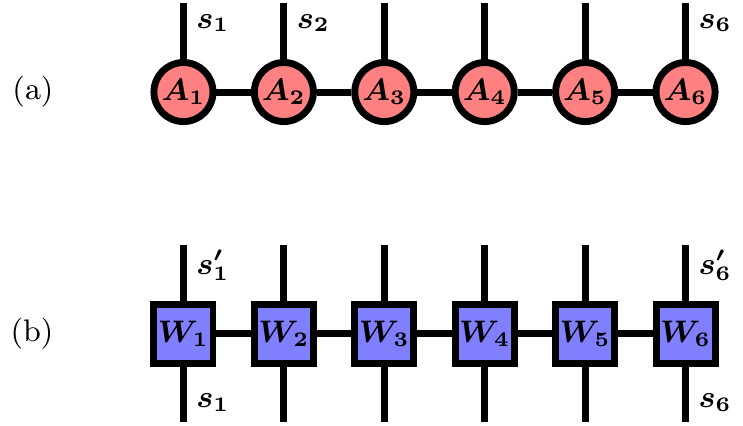}
   \caption{(a) Graphical representation of an MPS. Every circle
     corresponds to a tensor $A_i^{s_i}$ and each line to an index of
     this tensor. In this picture, the physical indices are the
     vertical lines, while the horizontal lines show the bond
     indices. Connected lines mean that the corresponding index is
     summed over. Fixing all the physical indices $s_i$ for each site
     results in a tensor of rank zero with the value of the coefficient
     $c_{s_1, \cdots, s_N} $. \\ 
   (b) Graphical representation of an MPO. The difference to an MPS is
   that an MPO has incoming indices $s_i$ and outgoing indices $s_i'$
   corresponding to the bra- and ket vectors of the operator. 
   }
   \label{fig:MPS}
 \end{figure}

In complete analogy to the states, one 
can factorize an operator into what is called a matrix-product
operator (MPO)~\cite{SchollwoeckDMRG_MPS},
\begin{equation}\label{eq:defMPO}
H = \sum_{\substack{ s_1, \cdots, s_N\\s_1', \cdots, s_N' }  }
W_1^{s_1, s_1'} \cdots W_N^{s_N, s_N'} \ket{s_1', \cdots, s_N'}
\bra{s_1, \cdots, s_N}, 
\end{equation}
where each $W_i^{s_i, s_i'}$ is a rank-4 tensor. Tensor networks in
general have a very useful graphical representation, which is shown
for an MPS and an MPO in Fig.~\ref{fig:MPS}. Note that when we use the
term MPS we {always} mean a one-dimensional chain of tensors as
shown in Fig.~\ref{fig:MPS}. 

To calculate Greens functions within the MPS formalism, one usually
first applies the 
DMRG~\cite{WhiteDMRG,SchollwoeckDMRG_MPS}, which acts on the
space of MPS and finds a variational ground state $\ket{\psi_0}$ and
ground-state energy $E_0$. It 
minimizes the expectation value 
\begin{equation}
	E_0 = \min_{\ket{\psi}} \frac{\braket{\psi | H | \psi  } } { \braket{\psi | \psi} }
\end{equation}
by updating usually two neighboring MPS tensors before moving on to
the next bond. This procedure also yields the Schmidt
decomposition of the state at the current bond on the fly. The DMRG
approximation is to keep only those states with the largest Schmidt coefficient. It is important to note that one can
perform a DMRG calculation for any
tensor network, as long as one can generate a Schmidt
decomposition~\cite{MurgVerstraeteTree}.  

For obtaining the Greens function, we employ an evolution in real time. Eq.~\eqref{eq:GF} is split into the
greater $G^>$ and lesser Greens function $G^<$: 
\begin{align}
 G(t)    &= -i \Theta(t) \bigg ( G^> (t) + G^< (t) \bigg) \nonumber \\
 G^> (t) &= \braket{ \psi_0 | c e^{-iHt} c^\dag | \psi_0} e^{ iE_0t}  \nonumber \\
 G^< (t) &= \braket{ \psi_0 | c^\dag e^{iHt} c | \psi_0} e^{-iE_0t} \text{,}
\end{align}
which are calculated in two separate time evolutions. This is done by
first applying $c^\dag$ (or $c$) and then time evolving this state and
calculating the overlap with the state at time $t=0$. 
The time evolution is the most computationally expensive part, 
since time evolved states are not ground states anymore, 
and the needed bond dimensions usually grow very fast with time. 

\subsection{\label{sec:FORKMPS}Fork Tensor Product States (FTPS)}

 \begin{figure}[t]
   \centering
   %\scalebox{.7}{ \input{tikzfigs/tree} }
   \scalebox{.7}{ \includegraphics{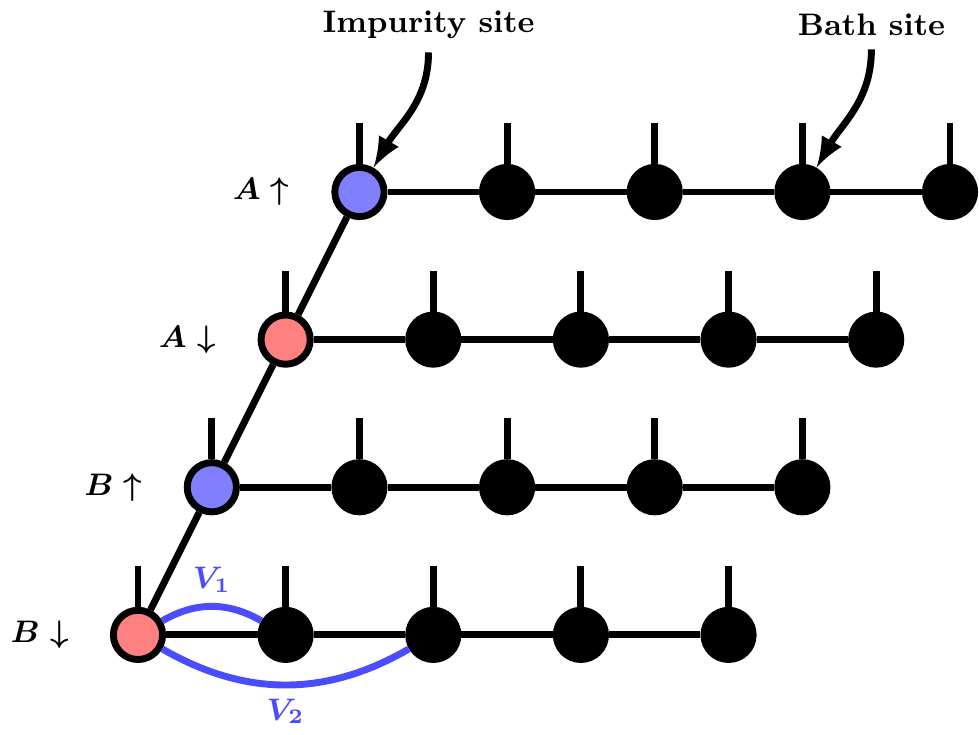} }
   \caption{Graphical representation of a fork tensor
     product state (FTPS) for multi-orbital AIM. The idea to separate bath degrees of
     freedom leads to a fork-like structure. In this picture, a
     two-orbital model with four bath sites each is shown. Orbitals are
     labeled $A$ and $B$ and the arrows denote the spin. Each
     spin-orbital combination has its \emph{own} bath sticking out to
     the right. As in Fig.~\ref{fig:MPS}, the vertical lines are the
     physical degrees of freedom (all of dimension two, for empty, resp.\ occupied bath sites). All other
     lines are bond indices and like in the MPS they are summed
     over. As mentioned in the text, the bath is represented in star
     geometry due to the smaller bond dimensions needed. 
    The bath sites are ordered according to their on-site energies.
    Two example
     hoppings $V_{1}$ and  $V_{2}$ are drawn.
   } 
   \label{fig:FORKMPS}
 \end{figure}

So far, the usual way of dealing with Hamiltonians like
Eq.~\eqref{eq:H_DMFT3B} using
MPS~\cite{Ganahl2BHubb,GanahlCheby,WolfCheby} has been to place the
impurity in the middle of the system and the up- and down-spin degrees of
freedom to its left and right, resp. The local state space of each
bath site then consists of $M$ spinless-fermion degrees of freedom, with
dimension $2^M$, where $M$ is the number of orbitals in the Hamiltonian
Eq.~\eqref{eq:H_DMFT3B}. This exponential growth is usually
accompanied by a very fast growth in bond dimension when using the above arrangement.
We did indeed encounter this very fast growth upon calculating the ground state
of some one- two- and three-orbital test cases.

For treatment by MPS, the general Hamiltonian
Eq.~\eqref{eq:H_DMFT3B} with hopping terms from the
impurity to each bath site 
is usually  transformed into a
Wilson chain  with nearest-neighbor hoppings only, i.e.\, of the
form $t_i(c_i^\dag c_{i+1}+\text{h.c.} )$~\cite{BullaNRG}. 
This was thought to be necessary
since long-range interactions look problematic for MPS-based
algorithms. Quite recently, though, it was discovered that MPS can deal
with the original form of $H_{\rm bath}$ in Eq.~\eqref{eq:H_DMFT3B}
better~\cite{WolfStarG}. Because all hopping terms in $H_{\rm bath}$ originate from the impurity,
this is called the \emph{star} geometry.
The reason for the better performance is that in the star geometry one has 
many nearly fully occupied (empty) bath sites with very low (high)
on-site energies $\epsilon_l$.

Since basis states with many unoccupied low-energy sites have a very low Schmidt coefficient, these states are discarded from the MPS.
The same holds for occupied high energy sites. However, when dealing
with multi-orbital models, the star geometry is not enough to
be able to calculate Greens functions using MPS. The growth
of the bond dimensions still makes those calculations unfeasible. 

The key idea of the present work is to construct a tensor network which is
beyond a standard MPS, but similar enough to be able to use
established methods like DMRG and time evolution. From Hamiltonian
\eqref{eq:H_DMFT3B} one can immediately notice that there are no terms
coupling bath sites of different orbitals. Hence, it might not be
advantageous to combine those, not directly interacting, degrees of
freedom into one large physical index in the MPS. 

Our proposed tensor network, therefore, separates the bath degrees of
freedom as 
much as possible. It consists of separate tensors for every
orbital-spin combination, each connected to bath tensors as shown in
Fig.~\ref{fig:FORKMPS}. This tensor network is no MPS anymore, since
there are some tensors (labeled $A\downarrow$ and $B\uparrow$ in the example of
Fig.~\ref{fig:FORKMPS}) that have three bond indices and one physical index, i.e.\ which are of rank 4. Cutting any bond
splits the network into two separate parts.
Therefore, one can calculate the Schmidt decomposition in a
way very similar to an MPS, which means that also DMRG is
possible. The main bottleneck of calculations with FTPS is to perform
SVDs of the rank-4 tensors representing the impurities. When all
bond indices have the same dimension $\chi$, it is necessary to do a SVD
for a $\chi^2d \times \chi$ matrix with computational complexity
$\mathcal{O}(\chi^4d)$. However, as we show below, this 
operation does not pose a substantial problem for calculations
using FTPS. Since the impurity tensors pose the biggest challenge, our tensor network would likely also
allow us to deal with the chain geometry without a drastic increase in computational cost. 
In the present paper we will only use FTPS with baths in star geometry.
\\
The proposed FTPS are similar to the tensor network used by Holzner
\emph{et al.}~\cite{HolznerForkMPS} to perform NRG
calculations for ground state properties of an AIM 
with two orbitals.

The three-legged tensors in our network (Fig.~\ref{fig:FORKMPS}) can
also be interpreted as two coupled junctions with three legs in the
language of Ref.~\cite{GuoWhiteYjunctions}, where it has been shown
that DMRG is possible on such junctions.
Furthermore, our approach has similarities with the so called Tree Tensor Networks (TTN)
\cite{ShiVidal_TTN,VTagliacozzoVidal_TTN_sim2Dsystems,
  MurgVerstraeteTree,PizzonrVerstraete_TTNentspectra}.

\subsubsection{Time Evolution}

Time evolution with the Hamiltonian Eq.~\eqref{eq:H_DMFT3B} is not
straightforward, since it features long-range hoppings. Possible
methods include Krylov approaches \cite{SchmitteckertKrylov}, the time-dependent variational
principle~\cite{HaegemanTDVP1,HaegemanTDVP2} and the series expansion
of $e^{iHt}$ proposed by Zaletel \emph{et
  al.}~\cite{ZaletelLongRange}. In this work, however, we use a much
simpler approach. 

First, we split the Hamiltonian into the following terms: 
(i) the spin-flip and 
pair-hopping terms $H^\text{SF-PH}_{m,m'}$ for
each orbital combination, with
$\sum_{m'>m}H^\text{SF-PH}_{m,m'}=H_{\text{SF-PH}}$ (see
Eq.~\eqref{eq:H_DMFT3B}), (ii) the density-density
interaction terms $H_{\text{DD}}$, and (iii) all other terms $H_{\text{free} } = H_{ \text{bath} } + \epsilon_0 \sum_{m \sigma} n_{m 0 \sigma} $. With these terms, we write the
time-evolution operator for a small time
step $\Delta t$ using a second-order Suzuki-Trotter
decomposition~\cite{SuzukiDecomp}, 

\begin{align}
  e^{-i \Delta t H } &\approx \left ( \prod_{m'>m} e^{-i \frac{ \Delta
        t } {2} H^\text{SF-PH}_{m,m'} } \right ) \cdot e^{ -i \frac{ \Delta
      t } {2} H_{\text{DD} }  } \cdot \nonumber \\  
  \times &e^{ -i \Delta t  H_{ \text{free} }  } \cdot e^{ -i \frac{
      \Delta t } {2} H_\text{DD}  } \cdot \left ( \prod_{m'>m}  e^{-i
      \frac{ \Delta t } {2} H^\text{SF-PH}_{m,m'} } \right ) \text{.} 
  \label{eq:timeEvol}
\end{align}
Note that in this decomposition, the order of the spin-flip and
pair-hopping terms is important. The order of operators in the
second product must be opposite to the one in the first.

We see that Eq.~\eqref{eq:timeEvol} involves three different operators
$H^\text{SF-PH}_{m,m'}$, $H_\text{DD}$ and $H_{ \text{free} }$, each of which
will be treated differently.

\emph{Time evolution of the density-density interactions} is performed
with an MPO-like representation of the time-evolution operator $e^{ -i
  \frac{ \Delta t } {2} H_\text{DD}  }$. For a three-orbital model, first the
full matrix ($4^3 \times 4^3$) of $e^{ -i \frac{ \Delta t } {2} H_{DD}
}$ is created, which is then decomposed into MPO-form by repeated
SVDs. Since $H_\text{DD}$ only consists of density-density interactions, no
fermionic sign appears in $e^{-i\Delta t H_\text{DD } }$. 

\emph{Time evolution of the spin-flip and pair-hopping terms} is more
involved than the density-density interactions, since the operators
change the particle numbers on the impurity sites. Therefore, it can be difficult
to deal with the fermionic sign of the time evolution
operator when the impurities are not next to each other in the
fermionic order. It turns out that the spin-flip and the pair-hopping
terms have the property $\hat{A}^3=\hat{A}$ individually, with
$\hat{A}$ being either the spin-flip or the pair-hopping
operator, resp. Furthermore they commute with each other allowing us to
separate them without Trotter error. The time-evolution operator
of $J\hat{A}$ is then given by:
\begin{equation}
 e^{-i \Delta t J \hat{A} } = \mathds{1} + \hat{A}^2 \big ( \cos
 \left( \Delta t J \right ) -1\big )- i \hat{A} \sin \left( \Delta t J
 \right ) \text{.} 
 \label{eq:SFTEVOMPO}
\end{equation}
For this operator, an MPO can be found for which the fermionic sign can
easily be determined\footnote{Note that we use the term MPO a bit 
  loosely here. What we mean is an operator factorized in the same
  fork-like structure as the state in Fig.~\ref{fig:FORKMPS}.
}.

\emph{To time evolve the bath terms} we use an iterative second-order
Suzuki-Trotter breakup for \emph{each} term in $H_{\text{free}
}$. Neglecting orbital ($m$) and spin ($\sigma$) indices, the first
step in this breakup is the following: $e^{-i \Delta t
  \sum_{l=1}^{N_b} H_l } 
\approx
 e^{-i \frac{\Delta t}{2} H_1 } \cdot e^{-i
  \Delta t \sum_{l=2}^{N_b} H_l } \cdot e^{-i \frac{\Delta t}{2} H_1
}$. Next we split off $H_2$ and iterate this procedure until we end up
with

\begin{align}\label{eq:BathTevo} 
  e^{ -i \Delta t  H_{\text{free}   } }  \approx \prod_{m\sigma} \Bigg
  [ \Bigg ( &\prod_{l=1}^{N_b-1} e^{ -i \frac{\Delta t}{2} H_{m l
      \sigma } } \Bigg ) \cdot   e^{ -i \Delta t H_{mN_b \sigma } }
  \nonumber \\ 
  & \cdot \Bigg ( \prod_{l=N_b-1}^{1}  e^{ -i \frac{\Delta t}{2}  H_{m
      l \sigma } } \Bigg )  \Bigg ] \text{,} 
\end{align}
with $N_b$ the number of bath sites and $H_{ml\sigma} = \epsilon_l
n_{ml\sigma} + V_l \left ( c_{m0\sigma}^{\dag} c_{ml\sigma} +
  \text{h.c.} \right )$. In the above equation, we neglected the term
  $\epsilon_0 n_{m0\sigma}$ that we add to $H_{m1\sigma}$. Eq.~\eqref{eq:BathTevo} 
  is a product of two-site gates (an operator acting non-trivially only on two sites)
with one of the two sites always being the impurity. This means that
those two sites are not nearest neighbors in the tensor network. To
overcome this problem, we use so called swap gates~\cite{SchollwoeckDMRG_MPS,StoudenmireMETTS}. The two-site
operator
\begin{equation}\label{eq:SwapGate}
  S_{ij} = \delta_{s_i,s_j'} \delta_{s_j,s_i'} \cdot (-1)^{n_i n_j}
\end{equation}
swaps the state of site $i$ ($s_i$ with occupation $n_i$) with the
state of site $j$ ($s_j$ with occupation $n_j$). The factor $(-1)^{n_i
  n_j}$ gives the correct fermionic sign and is negative if an odd
number of particles on site $i$ gets swapped with an odd number of
particles on site $j$. To be more precise, the matrix representation of
the swap gates used in this work is: 
\begin{equation}\label{eq:SwapGateExpl}
  S_{ij} = \ket{00} \bra{00} + \ket{10} \bra{01} +\ket{01} \bra{10} -
  \ket{11} \bra{11}  \text{.} 
\end{equation}
It turns out that every swap gate can be combined with an actual time
evolution gate without additional computational time. For example, the
first step in this time evolution would be to apply $e^{-i\frac{\Delta
    t}{2} H_{m1\sigma} }$. Immediately afterwards, even before the SVD
(to separate the tensors again), the swap gate is applied so that the
impurity and the first bath sites are swapped. By repeating this process
one moves the impurity along its horizontal arm in
Fig.~\ref{fig:FORKMPS}. Because a second-order decomposition is used,
now all time evolution gates except the one at site $N_b$ have
to be applied again. But now, the impurity and bath site needs to be swapped
\emph{before} time evolution. 
\\ \\
Note that the algorithm presented above cannot
only be used to perform real-time evolutions, but it is applicable also to
evolution in imaginary time simply by replacing $idt$ by $d\tau$.

\section{\label{sec:DMFT}Multi orbital DMFT with FTPS}
In this section we present details of our impurity solver.

We refer to Refs.~\cite{GeorgesDMFT_easy,GeorgesDMFT}
for DMFT in general, and to Refs.~\cite{Lechermann_DMFTwithWannier, Kotliar_ElStrCal_DMFT} 
for DMFT in the context of realistic
ab-initio calculations for correlated materials. 

In the latter approach, called density-functional theory (DFT)+DMFT,
the correlated subspace is described by a 
Hubbard-like Hamiltonian. Within DMFT, this model is mapped onto
the AIM Hamiltonian \eqref{eq:H_DMFT3B}. This mapping defines the bath
hybridization function $\Delta(\omega)$ describing the influence of
the surrounding electrons. 

Since FTPS provide the Greens function of the AIM on the real-frequency axis,
also the self-consistency loop is performed directly for real
frequencies. For calculating the bath hybridization, we use retarded
Greens functions with a finite broadening $\eta_{SC}$ in order to avoid
numerical difficulties with the poles of the Greens function. Throughout
this work, we use $\eta_{SC}=0.005$\,eV\footnote{For stability reasons, a larger
  broadening of $\eta_{SC}=0.01$\,eV was used in the first two DMFT-cycles.}.

The impurity solver calculates the self energy $\Sigma(\omega)$ of
the AIM, given the bath hybridization function $\Delta(\omega)$ and the
interaction Hamiltonian on the impurity. To this
end, our solver performs the following steps, which are explained in
more detail in the text below: 
\begin{enumerate}
 \item \label{enum:DMFT2} Obtain bath parameters $\epsilon_{l}$ and
   $V_l$ by a deterministic approach based on integration of the bath 
hybridization function $\Delta(\omega)$. 
 \item \label{enum:DMFT3} Calculate the ground state $\ket{\psi_0}$
   and ground-state energy $E_0$ of the interacting problem. 
 \item \label{enum:DMFT4} Apply impurity creation or annihilation
   operators, and time evolve these states to determine the interacting
   Greens function ( Eq.~\eqref{eq:GF} ). 
 \item \label{enum:DMFT5} Fourier transform Eq.~\eqref{eq:GF} to obtain
   $G(\omega)$ and calculate the local self-energy, 
   \begin{equation}
     \Sigma (\omega) = G_0(\omega)^{-1} - G(\omega)^{-1} \text{.}
   \end{equation}
\end{enumerate}

To perform step \ref{enum:DMFT2} we use
\begin{align}
 \label{eq:BathDiscr}
 V_l^2 &= \int_{I_l} \left [ -\frac{1}{\pi}  \operatorname{Im}
   \Delta(\omega)  \right ]  d\omega \nonumber \\ 
 \epsilon_l &= \frac{1}{V_l^2} \int_{I_l} \omega \left [
   -\frac{1}{\pi} \operatorname{Im}  \Delta(\omega)  \right ] d\omega
 \text{,} 
\end{align}
similar to Refs.~\cite{BullaNRG} (NRG) and~\cite{WolfStarG}. Each interval 
$I_l$ corresponds
to a bath site. This discretization can be interpreted as representing each interval $I_l$ 
as a delta peak at position $\epsilon_l$ and weight $V_l^2$. Sum rules for such discretization parameters 
can be found analytically~\cite{KochSumRules}. In this work we choose the length of each interval such that the
area of the bath spectral function $-\frac{1}{\pi}  \operatorname{Im}
\Delta(\omega) $ is approximately constant for each
interval~\cite{deVegaDiscBath}. For the case at hand, this discretization was found to be
numerically more stable than using intervals of constant
length. Unless stated otherwise, we use $N_{b}=109$ bath sites per
orbital and spin. We note that this scheme is not restricted 
to diagonal hybridizations. In the general case of off-diagonal hybridizations the 
hybridization function is a matrix $\underline{\Delta}$. Therefore, instead of 
taking the imaginary part we can use the bath spectral function 
$ \frac{i}{2\pi}(\underline{\Delta} - \underline{\Delta}^\dag)$.
Similarly to Eq.~\eqref{eq:BathDiscr}, we represent each interval 
by one delta-peak for each orbital. For instance, fixing $\epsilon_l$ to the center of the interval,
the hopping parameters $V_l$ can be found systematically from the Cholesky factorization 
of $\int_{I_l}  \frac{i}{2\pi}(\underline{\Delta} - \underline{\Delta}^\dag) d\omega$.
Most importantly, this scheme does not involve any fitting procedure on the Matsubara 
axis. A very similar approach was developed independently in Ref.~\cite{Liu_NonDiagBath}.

In step~\ref{enum:DMFT3} we use a DMRG approach with the following
parameters, unless specified otherwise. The truncated weight $t_w$
(sum of all discarded singular values of each SVD) is kept smaller than
$10^{-8}$. When spin-flip and pair-hopping terms are neglected,
we use an even smaller cutoff of $10^{-9}$.
Note that, except in the five-band calculation, we do \emph{not} restrict the bond dimensions by some hard
cutoff (see Appendix~\ref{ssec:TrW}). 

During time evolution (step \ref{enum:DMFT4}), we use a truncated
weight of $t_w=2\cdot10^{-8}$, or $10^{-8}$ with density-density
interactions only.
We time evolve to $t=16$\,eV$^{-1}$,
with a time step of $\Delta t=0.01$\,eV$^{-1}$. Greens functions
are measured every fifth time step. The time-evolution operator of
$H_{\text{loc} }$ is applied using the zip-up algorithm
\cite{StoudenmireMETTS}. Afterwards the Greens functions are
extrapolated in time using the linear prediction
method~\cite{WhiteLinPred,Ganahl2BHubb} up to $t=250$\,eV$^{-1}$. 
Time evolution is split into two runs one forward and one
backwards in time \cite{BarthelTwoSidedTevo} to be able to reach longer
times.

In the Fourier transform to $\omega$-space (step \ref{enum:DMFT5}), we
use a broadening in the kernel $e^{i\omega t - \eta_{FT} |t| }$ of
$\eta_{FT}=0.02$\,eV to avoid 
cutoff effects remaining after the linear prediction. 
The influence of the linear prediction on our results is
discussed in Appendix~\ref{ssec:LP}. We want to stress that although
a calculation with full rotational symmetry is more demanding, the
computational effort is still very feasible. With the parameters
mentioned above one full DMFT-cycle takes about five hours on 16
cores. 

To verify that our implementation of DMRG and time evolution produces
correct results when used with our tensor network, we first compared
Greens functions and ground-state energies for $U=J=0$ for several
bath parameter sets. The next step of our testing was to include
density-density interactions, one term at a time. For example,
we only included $(U'-J)n_{10\uparrow}n_{30\uparrow}$ and compared
energy and Greens function to a standard one-orbital
MPS solver. Finally, we also compared our method to the MPS two-band
solver used in Ref.~\cite{Ganahl2BHubb}. Indeed all tests performed
produced correct energies and Greens functions.

\section{\label{sec:Results} Results}

We performed  DMFT calculations based on a band structure
obtained from density functional theory (DFT) for the prototypical
compound \ce{SrVO3}, using the approximation of the Kanamori Hamiltonian (Eq. \eqref{eq:H_DMFT3B}).
It has a cubic crystal structure with
a nominal filling of one electron in the V-3d shell
\footnote{ Indeed, model calculations done
for fillings of $N=2$ and $N=3$ electrons, the latter in the insulating phase, show that these
calculations are of comparable computational effort. For any $N$ we do 
expect increased numerical effort close to the Mott transition though.}. Due to 
the
crystal symmetry, the five orbitals of the V-3d shell split into two
$e_g$ and three $t_{2g}$ orbitals. The latter form the correlated
subspace. We performed the DFT calculation with Wien2k~\cite{Wien2k},
and used 34220 k-points in the irreducible Brillouin
zone in order to reach an energy resolution comparable with the
$\eta_{SC}=0.005$\,eV broadening. 

\begin{figure}[t]
   \centering
   %\scalebox{1}{ \input{tikzfigs/SF} }
   \includegraphics{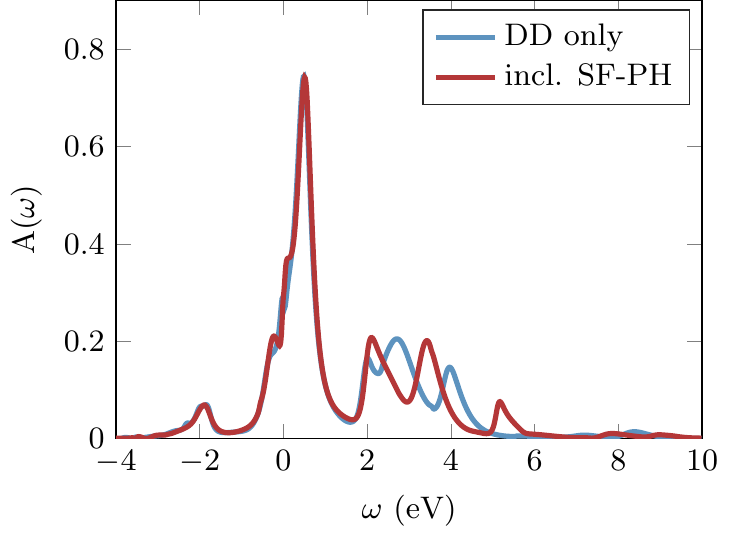}
   \caption{Spectral functions $A(\omega)$ for density-density
     interactions (DD) only (blue line), and with spin-flip and
     pair-hopping terms included (red line). In both calculations 
     we used $U=4.0$\,eV and $J=0.6$\,eV. Both spectra show a 
     three-peak structure in the upper Hubbard band and additional 
     features at high energies (around $8$\,eV).} 
   \label{fig:SFterms}
  \end{figure}  

   \begin{figure}[t]
   \centering
   %\scalebox{1}{ \input{tikzfigs/ReconDeltaNbs} }
   \includegraphics{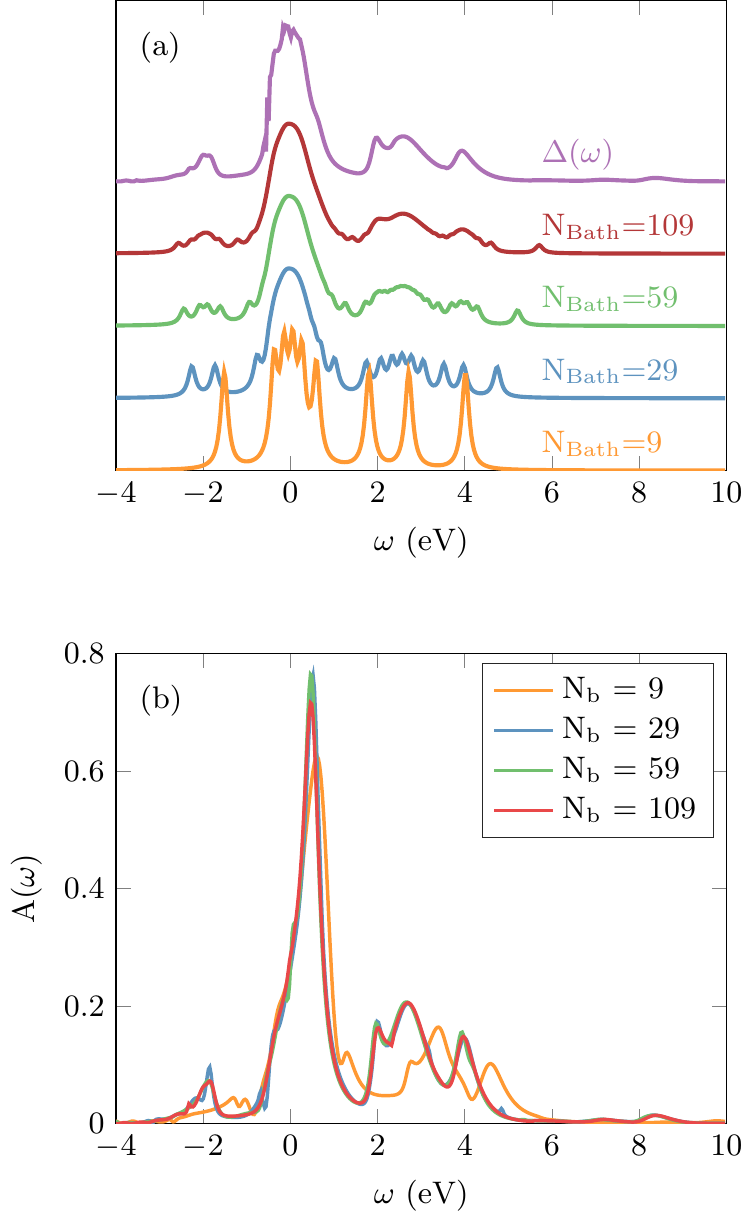}
   \caption{(a) We take the bath spectral function $\Delta(\omega)$ from
     the DMFT self-consistent solution for $N_b=109$ and represent it
     using various numbers of bath sites. 
     It is obvious that $N_b=9$ is too small to represent the
     bath well.
   (b) Converged DMFT spectral function using the AIM with
   different numbers of bath sites. Only the smallest bath shows a
   noticeable difference. This is mostly due to the fact that in this case a 
   higher broadening of
   $\eta_{FT}=0.1$\,eV had to be used in the Fourier transform and time evolution was only
   possible to $t=14\text{ eV}^{-1}$. The additional small structure
   at $\omega=0$ for $N_b=59$ bath sites is most likely a linear prediction artifact.
 }
   \label{fig:diffBathSizes}
 \end{figure}

The TRIQS/DFTTools package
(v1.4)~\cite{TRIQS_DFTTOOLS,TRIQS_DFTTOOLS_1,TRIQS_DFTTOOLS_2}, which is based on the
TRIQS library (v1.4)~\cite{TRIQS}, was used to generate the projective
Wannier functions and to perform the DMFT self-consistency cycle.

Fig. \ref{fig:SFterms} shows the main results of this paper, the DMFT
spectral function $A(\omega)$ for \ce{SrVO3}, 
(i) in the approximation of density-density interactions only
and (ii) with full rotational invariance including spin-flip and pair-hopping
terms. Overall, both cases show the well known features of the
\ce{SrVO3} spectral function~\cite{SVO_DMFT1,SVO_DMFT2}. We see a hole excitation at around
$-2$\,eV, and the quasi-particle peak at zero energy whose shape and
position does not depend on the inclusion of full rotational invariance. In
the upper Hubbard band, a distinctive three-peak structure can
be seen. This structure has not been resolved in other exact methods like
CTQMC (problem with analytic continuation, see below) or NRG 
(logarithmic discretization problem). In our real time approach, high energies 
correspond to short times, where the calculations are particularly 
precise\footnote{We note that high energy peaks already appear in the first 
DMFT iteration, for which the bath does not have any spectral weight at 
high energies.}. Most methods allow
to resolve structures in the Hubbard bands only in special cases (see
Ref. \cite{SangiovanniPeakStructHubbardTJ} for an example using ED). Of course,
atomic-limit based algorithms such as the Hubbard-I approximation or
non-crossing approximation (NCA) show
atomic-like features, but they have very limited accuracy for the description of the
low-energy quasi-particle excitations in the metallic phase~\cite{GullNCA}. Thus, our
FTPS solver combines the best of the two worlds, with atomic multiplets
at high energy and excellent low-energy resolution \emph{at the same time}.

The energies of the three peaks in the upper Hubbard band differ depending on whether SF-PP
terms are taken into account or not. Details of this peak structure, as well as additional 
excitations visible at higher energies, will be discussed below in Sec.~\ref{ssec:discPeaks}.  

First we examine the convergence of our results with respect to the number 
of bath sites and compare our spectrum to CTQMC.
The following discussion is mostly based on
calculations without spin-flip and pair-hopping terms. In this case,
the calculations can be done faster and with higher precision, since
there is no particle exchange between impurities. In all subsequent 
plots, we show results from calculations with DD interactions only.

\subsection{\label{ssec:BathSize}Effect of Bath Size}

In order to achieve a reliable high resolution spectrum on the real-frequency axis, it is
imperative to have a good representation of the hybridization
function $\Delta(\omega)$ in terms of the bath parameters,
for which a sufficient number of bath sites is
needed. Fig.~\ref{fig:diffBathSizes} shows how well a hybridization
function can be represented with our approach (Eq. \eqref{eq:BathDiscr})
using a certain number of bath sites. We see that for $N_b= 9$ bath sites
(we always denote sites per orbital), $\Delta(\omega)$ can be
reconstructed only very roughly, which in turn gives an incorrect spectral
function (Fig.~\ref{fig:diffBathSizes} bottom). To some extent, the difference
in the spectrum is due to the shorter time evolution and therefore a
higher broadening $\eta_{FT}$ we were forced to use. For such a small
bath, the finite size effects from reflections at the bath ends 
appear much earlier in the time evolution.

Increasing the number of bath sites to $N_b=29$, we observe
that the reconstructed bath spectral function already shows the relevant
features of $\Delta(\omega)$. 
The spectrum is well converged for the largest bath sizes
$N_b=59$ and $N_b=109$.

 \begin{figure}[ht]
   \centering
   %\scalebox{1}{ \input{tikzfigs/QMCvsTPS} }
   \includegraphics{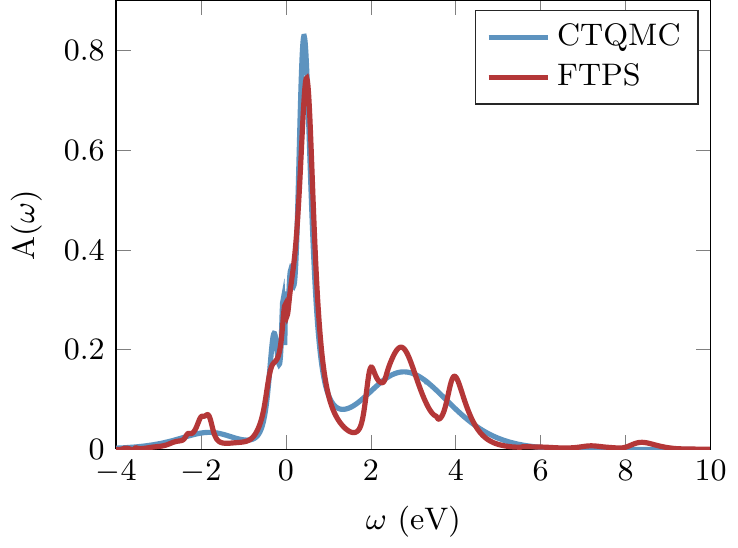}
   \caption{DMFT spectral functions $A(\omega)$ from CTQMC+MaxEnt (blue line)
     at $\beta=200$ eV$^{-1}$, and from FTPS (red line). The FTPS result
     shows a distinctive three-peaked structure in the upper Hubbard
     band.} 
   \label{fig:A}
 \end{figure}
 
  \begin{figure}[ht]
   \centering
   %\scalebox{1}{ \input{tikzfigs/Gtau} }
   \includegraphics{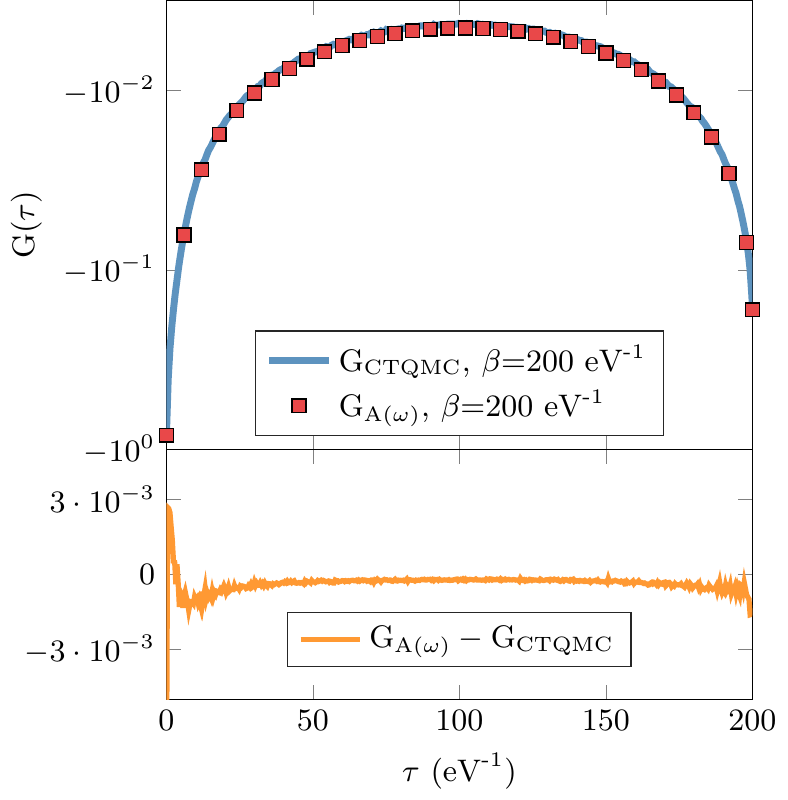}
   \caption{Comparison of imaginary time Greens functions $G(\tau)$
     from CTQMC (G$_{\text{CTQMC}  }$, blue line) and FTPS using Eq.~\eqref{eq:AOmToGtau} 
     ($\text{G}_{\text{A}(\omega) }$, red
     squares). The agreement is equally good also at $\beta=100$~eV$^{-1}$ 
     and $\beta=400$~eV$^{-1}$ (not shown). The difference between the two 
Greens functions is shown in the bottom panel. Note that on both ends 
$\text{G}_{\text{A}(\omega) }$ is smaller than G$_{\text{CTQMC}}$. The 
normalization of the spectral function demands that $G(\tau=0)+G(\tau=\beta) = 
-1$. The CTQMC data deviates in the order of $10^{-2}$ from this constraint due 
to statistical noise, while FTPS gives (by construction) the correct result to 
a precision of $10^{-8}$. This explains the bigger differences of the Greens 
functions around $\tau=0$ and $\tau=\beta$. For better visibility of the 
$\tau>0$ data, the value of $9\cdot10^{-3}$ at $\tau=0$ is not shown. } 
   \label{fig:Gtau} 
 \end{figure}
 
   \begin{figure}[ht]
   \centering
   %\scalebox{1}{ \input{tikzfigs/AnaCont} }
   \includegraphics{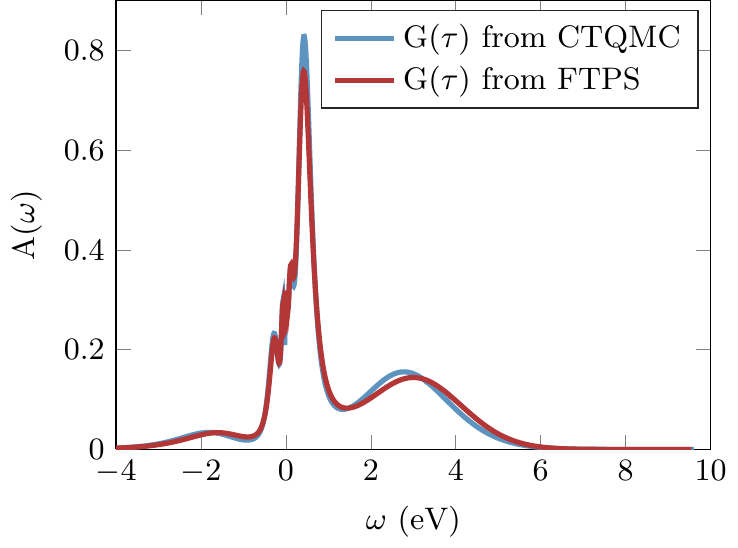}
   \caption{Spectral functions from analytically continued
     imaginary-time Greens functions $G(\tau)$ calculated by CTQMC (blue line) 
     and by FTPS (red line). Clearly, the analytic continuation cannot resolve 
     the peak structure in the upper Hubbard band.} 
   \label{fig:AnaCont}
 \end{figure}

\subsection{\label{ssec:CompareQMC}Comparison to CTQMC} 

In Fig.~\ref{fig:A} we compare the converged spectral function of our
approach (FTPS) with a spectrum obtained from CTQMC and analytic continuation. 
In both calculations, we used the same interaction Hamiltonian with
density-density interactions only. The CTQMC calculation was performed with the TRIQS 
CTHYB-solver (v1.4)~\cite{TRIQS_CTHYB, WernerMillisHybExpCTQMC} with
$3.2\cdot10^{7}$ measurements and at inverse temperature $\beta=200$\,eV$^{-1}$. 
For the analytic continuation we applied the $\Omega$MaxEnt method~\cite{OMEGA_MaxEnt}.
 
The three-peak structure in the upper Hubbard band is
not present in the CTQMC spectrum. We will show below in an example that even 
for a Greens function that does contain these peaks the analytic continuation 
does not resolve this structure.

For another comparison, we consider the imaginary time 
Greens functions $G(\tau)$ in Fig.~\ref{fig:Gtau}. Apart from the effect of 
statistical errors, CTQMC provides an exact self consistent solution of DMFT on 
the imaginary-frequency axis. As mentioned above, when we use the FTPS solver, 
we formulate the DMFT self-consistency equations on the real-axis. To obtain an 
approximate finite temperature imaginary-time Greens function from FTPS that we 
can compare to the CTQMC result, we need to take the finite temperature of the 
CTQMC calculation into account. Therefore, we use the FTPS spectrum 
$A(\omega)$ and assume that we would obtain the same spectrum for a finite (but 
high enough) inverse temperature $\beta$, and use: 

\begin{equation} 
\label{eq:AOmToGtau}
 G(\tau) = \int \frac{d\omega}{2\pi} A(\omega) \frac{ e^{-\omega \tau}
 }{ e^{-\beta \omega}+1 } \text{} 
\end{equation}
The results in Fig.~\ref{fig:Gtau} show very good agreement on a 
logarithmic scale.

Another important indication of the validity of our results is the
value of $A(\omega=0)$. To get a comparable number, we use the CTQMC
imaginary time Greens function $G(\tau)$
and Fourier transform it to get $G(i\omega_n)$: 
\begin{align*}
 G(i\omega_n) = \int  e^{i\omega_n \tau} G(\tau) d\tau \text{.}
\end{align*}
Looking at the last few DMFT-cycles, we estimate it to be around
$A(\omega=0) = -\frac{1}{\pi} \lim_{i\omega_n \to 0}\Im G(i\omega_n) \approx 0.272$\,eV$^{-1}$ 
with fluctuations in the last digit.
For the FTPS, the exact height of $A(\omega=0)$ of the
FTPS spectrum changes a little for each DMFT iteration,
mainly due to slight variations in the linear prediction.
Using the same prescription as for CTQMC, we estimate
it to be $A(\omega=0) = 0.28 \text{ eV$^{-1}$,}$
with fluctuations of about $0.01 \text{ eV$^{-1}$}$.
This agreement is very good considering that linear prediction has its
strongest influence at small energies. Further benchmarks
concerning the linear prediction can be found in Appendix~\ref{ssec:LP}.

Finally, we show that the ill-posedness of the analytic continuation
is the most likely explanation for the missing peak structure in the upper 
Hubbard band of the spectral function obtained from the CTQMC data.
To do so, we take the FTPS spectrum $A(\omega)$, calculate $G(\tau)$
as described above, and perform the same
analytic continuation that we did for the $G(\tau)$ from CTQMC. We
added noise of the order of 
the CTQMC error to the FTPS data. The resulting spectrum is shown in
Fig.~\ref{fig:AnaCont}, and indeed the peak structure in the
upper Hubbard band vanishes.

   \begin{figure}[ht]
   %\scalebox{1}{ \input{tikzfigs/Js} }
   \includegraphics{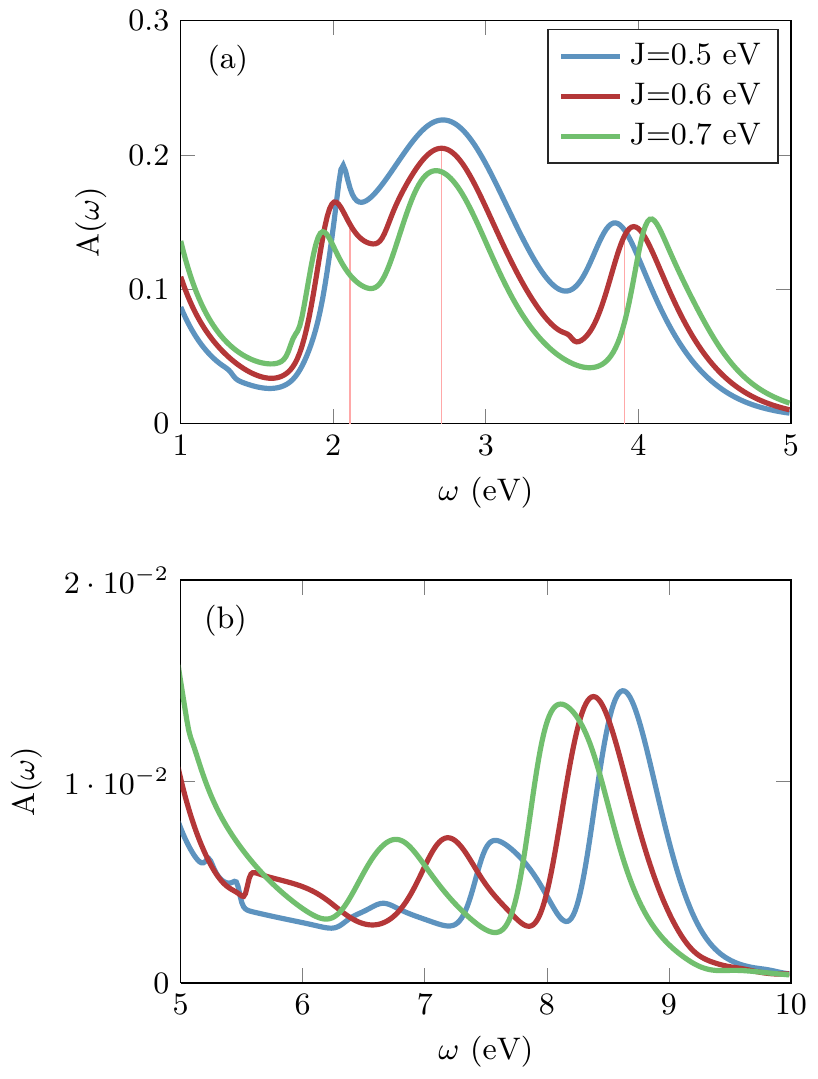}
   \caption{(a) Closeup of the three-peak structure for various
     values of $J$. Additionally, we show vertical lines for the $J=0.6$\,eV
     spectrum at energies $\omega_M$ (position of the middle peak) and
     at $\omega_M+2J$ and $\omega_M-J$. We see that the width of the
     upper Hubbard band is close to $3J$. 
   (b) Closeup of the small spectral peaks at high
   energies. These correspond to excitations into the $N=3$ sector of
   the atomic model (see Tab.~\ref{tab:atomiclevels}). The height
   of each peak can be estimated by the degeneracy of the atomic
   states. Effective parameters $\bar{J}$ are $0.53$\,eV ($J=0.5$\,eV),
   $0.59$\,eV ($J=0.6$\,eV) and $0.68$\,eV ($J=0.7$\,eV). They are obtained from the 
   difference between the two peaks highest in energy.
   } 
   \label{fig:J07}
 \end{figure}
 
 \subsection{\label{ssec:discPeaks} Discussion of Peak Structure -
   Effective Atomic Physics} 

 In order to understand the peak structure observed in the spectral
 functions, we take a look at the underlying atomic problem, where for
 simplicity we start with density-density interactions only. 
 We will show that the same arguments hold for full rotationally invariant
 interactions.

 Tab.~\ref{tab:atomiclevels} shows the relevant atomic states and their
 corresponding energies. The atomic model has a hole excitation at
 energy $-\epsilon_0$ and three single electron excitations with energies
 $U+\epsilon_0$, $U-2J+\epsilon_0$ and $U-3J+\epsilon_0$ relative to
 the ground state. If we measure the energy differences between the three
 peaks of the upper Hubbard band 
 in our results, we find values of $1.27$\,eV
 and $0.69$\,eV, which is close to the atomic energy differences
 of $1.2$\,eV and $0.6$\,eV ($J=0.6$\,eV). We also find the hole
 excitation at $-2.0$\,eV. This indicates that we can describe the positions
 of the observed peaks approximately by atomic physics with effective
 parameters $\bar{\epsilon}_0$, $\bar{U}$ and $\bar{J}$ and widened peaks. 
 Furthermore, the heights of the peaks roughly correspond to
 the degeneracy of the states in the atomic model (see Tab.~\ref{tab:atomiclevels}).

\begin{table*}[t]
   \caption{\label{tab:atomiclevels} Relevant states of the atomic
     problem of Hamiltonian \eqref{eq:H_DMFT3B} without spin-flip and
     pair-hopping terms.
} 
  \begin{ruledtabular}
  \begin{tabular}{ lccc }

    type 		& states & energy difference to ground state & degeneracy \\ \hline
    $N=1$, ground state	& $\ket{\uparrow,0,0} \text{ } \ket{\downarrow,0,0} \text{ } \ket{0,\uparrow,0} \cdots $  & $0$  & 6\\ \hline
    $N=0$ &  $\ket{0,0,0}$ & $-\epsilon_0$ & 1 \\   \hline
    $N=2$, same spin    	& $\ket{\uparrow,\uparrow,0} \text{ } \ket{\uparrow,0,\uparrow} \text{ } \ket{0,\uparrow,\uparrow} \cdots$ & $U-3J+\epsilon_0$ & 6 \\
     $N=2$, different spin  	& $\ket{\uparrow,\downarrow,0} \text{ } \ket{\uparrow,0,\downarrow} \text{ } \ket{\downarrow,\uparrow,0} \cdots$ & $U-2J+\epsilon_0$ & 6 \\
    $N=2$, double occupation	& $\ket{\uparrow \downarrow,0,0} \text{ } \ket{0, \uparrow \downarrow,0} \text{ } \ket{0,0,\uparrow \downarrow} $   & $U+\epsilon_0$ &3 \\     \hline

    $N=3$, all spins equal    	& $\ket{\uparrow ,\uparrow,\uparrow} \text{ } \ket{\downarrow,\downarrow,\downarrow} $ & $3U-9J+2\epsilon_0$ & 2 \\
    $N=3$, one spin different   	& $\ket{\uparrow ,\uparrow,\downarrow} \text{ } \ket{\uparrow ,\downarrow,\uparrow} \text{ } \ket{\downarrow,\uparrow, \uparrow} \cdots $ & $3U-7J+2\epsilon_0$ & 6 \\
    $N=3$, double occupation    	& $\ket{\uparrow \downarrow ,\uparrow,0} \text{ } \ket{\uparrow \downarrow,\downarrow,0} \text{ } \ket{\uparrow \downarrow,0,\uparrow} \cdots$ & $3U-5J+2\epsilon_0$ & 12 \\

   \end{tabular}
  \end{ruledtabular}
\end{table*}

We can determine $\bar{U}=5.97$\,eV (where $U=4.00$\,eV) from the energy
difference of the peak highest in energy to the hole excitation. 
This increase of $\bar{U}$ compared to $U$ is plausible considering the following. 
When coupling the impurity to the bath, particles have the possibility to
avoid each other by jumping into unoccupied sites of the bath. This
results in a decrease of $\langle n_{\uparrow} n_{\downarrow}
\rangle$. To model this situation using atomic physics, one needs to
increase the interaction strength. Finally, it is well known that $J$
is much less affected by the surrounding electrons than $U$,
since the latter is screened significantly stronger~\cite{VaugierScreening}. 

Tab.~\ref{tab:rescParam} shows how bare atomic parameters change when
adding a bath and we see that our qualitative arguments give a correct
idea of how parameters are rescaled.

\begin{table}[t]
   \caption{\label{tab:rescParam} Atomic parameters and their effective
     values obtained from the spectral functions shown in
     Fig.~\ref{fig:A} and \ref{fig:J07}. For $J$ the values itself were obtained
     from the energy difference of the highest peak to the lowest peak, whereas the 
     uncertainty is estimated from $\omega_M+2J$ and $\omega_M-J$.
} 
 \begin{ruledtabular}
  \begin{tabular}{ ccc }
    parameter 		& atomic value (eV) 	& effective value (eV) 	\\ \hline
    $\epsilon_0$	& -$0.86$		& -$2.00$		\\
    $U$			& $4.00$		&  $5.97$		\\
    $J$			& $0.50$		&  $0.59(6)$		\\
    $J$			& $0.60$		&  $0.66(3)$		\\
    $J$			& $0.70$		&  $0.72(2)$		\\
   \end{tabular}
  \end{ruledtabular}
\end{table}
Further evidence that the observed three-peaked structure is indeed a
result of atomic physics can be seen in Fig.~\ref{fig:J07}. It shows a
closeup of the upper Hubbard band for three different values of
$J$. The corresponding effective parameters $\bar{J}$ are shown in Tab.~\ref{tab:rescParam}. 
We observe that also $J$ is rescaled slightly,
but the rescaling gets smaller for higher $J$. 
Furthermore, increasing $J$ also increases the total width of the
Hubbard band, which scales mostly linearly with $J$. At the same time, 
measuring the quasi-particle spectral weight as a function of $J$ at
constant $U$ shows that it \emph{increases} with increasing $J$,
implying also an increasing critical $U_c$ for the metal-to-insulator
transition~\cite{GeorgesHundJ}.

Upon a careful inspection of the spectral function in Fig.~\ref{fig:A},
we observe small peaks at energies around $8$\,eV. A closeup of this
energy region for different values of $J$ is shown in
Fig.~\ref{fig:J07}. The energy difference between the peaks is close to
$2J$ and can, again, be well explained by atomic physics, namely excitations into states
with 3 electrons on the impurity (Tab.~\ref{tab:atomiclevels}) \footnote{Nevertheless, the
effective parameters $\bar{J}$ differ a little from those obtained from the main
Hubbard band.}. These excitations originate from small admixtures of $N=2$ 
states to the ground state. 

With atomic physics in mind, let us take a look again at the spectrum
using full rotational symmetry (Fig.~\ref{fig:SFterms}). The 
spin-flip and pair-hopping terms only contribute if there are two or
more particles present. Thus, the quasi-particle peak and the hole
excitation do not change. The atomic $N=2$ sector does change,
however. Diagonalizing the Hamiltonian, we find eigenstates with
three different energies and differences of $3J=1.8$\,eV and $2J=1.2$\,eV,
resp. Measuring the energy differences in
Fig.~\ref{fig:SFterms}, we find $3\bar{J} = 1.75$\,eV and $2\bar{J} =
1.32$\,eV. Estimating $\bar{U} = 5.81(5)$ we see that it does not change much 
compared to DD only\footnote{Note that the peak highest in energy has an atomic 
energy of $E=U+2J+2\epsilon_0$. Therefore, $\bar{U}$ can only be determined after 
$\bar{J}$ is found.}. Again, we can describe the spectrum approximately by atomic physics with
effective parameters. Like in the case with density-density terms only, we also see
the tiny excitations to states belonging to the atomic $N=3$ sector.

\subsection{\label{ssec:BeyondAtomic} Beyond Atomic Physics  }

\begin{figure}[t]
   \centering
   %\scalebox{1}{ \input{tikzfigs/Us} }
   \includegraphics{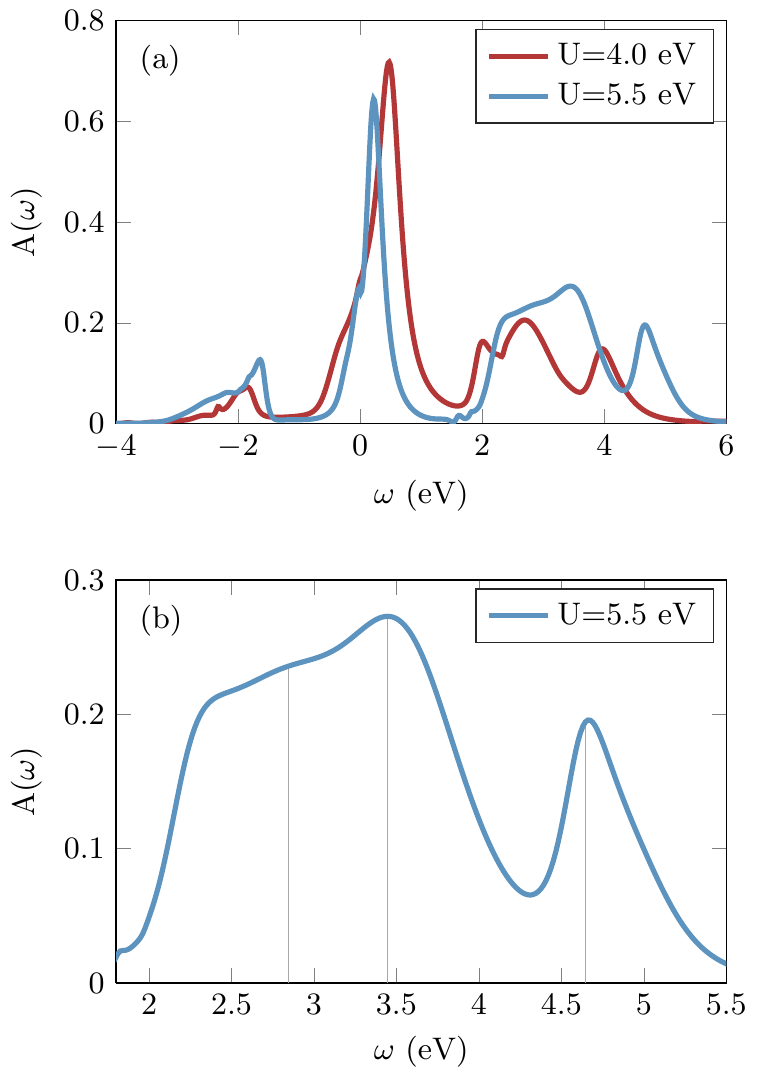}
   \caption{(a) Increasing $U$ results in a slimmer central
     peak and a shift of the upper Hubbard band. Also the
     three-peaked structure gets smeared out. 
   (b) Closeup of the upper Hubbard band. As in
   Fig.~\ref{fig:J07}, additional vertical lines are plotted at
   $\omega_{M}$ (position of the middle peak) and at $\omega_M + 2J$
   and $\omega_M - J$ as a rough guide to where the atomic peaks would
   be located. With the help of these lines one can discern a three-peaked
   structure again, but extended by a feature at the inside of the
   Hubbard band.  } 
   \label{fig:Us}
  \end{figure}

The previous section showed that at $U=4.0$\,eV the spectral features in the
Hubbard bands can be well described by atomic physics
with effective parameters and widened peaks. It is not clear whether this picture
is valid for higher interaction strengths $U$ in the metallic regime. 
In Fig.~\ref{fig:Us} we show results with $U=5.5$\,eV at constant $J=0.6$\,eV. We see a
shift of the upper Hubbard band to higher energies, but little shift
of the hole-excitation. Also the central peak is shifted and gets slimmer since more
weight is transferred into the Hubbard bands.  
Most importantly, as
we approach the strongly-correlated metallic regime, we
clearly leave the realm where atomic physics can describe all the
spectral features.  

We find that the three-peak structure of the upper
Hubbard bands smears out, and even
vanishes. The closeup of the upper Hubbard band in
Fig.~\ref{fig:Us} shows that with the help of the bare energy
differences all three atomic peaks can be discerned again, accompanied by an
additional structure at the low-energy side of the Hubbard band,
which is reminiscent of the Hubbard side peaks
found in the one- and two-band Hubbard model on the Bethe lattice~\cite{Ganahl2BHubb} upon increasing $U$. 
We leave further investigation of this feature to future work.

It might at first seem counter-intuitive that increasing $U$ makes the
physics less atomic-like. Indeed, at very high interaction strengths,
in the insulating regime, the spectrum must become atomic-like again. 
Here, however, we identify an intermediate regime where additional
structures appear when increasing $U$, since we get closer to the Mott
metal-to-insulator transition.  
\vspace{2cm}

\subsection{Solution of a five-band AIM}
  In this section we show that FTPS can not only deal 
  with three-band models, but also work in the case of five orbitals. 
  To do so, we use the bath parameters $\epsilon_k$ and $V_k$ from 
  the converged $N_b=59$ calculation for \ce{SrVO3} and construct an 
  artificial degenerate five-band AIM. Interaction 
  parameters are $U=4.0$\,eV and $J=0.6$\,eV. We decrease the on-site 
  energy $\epsilon_0$ to get a similar occupation of each impurity 
  orbital as for \ce{SrVO3} $(\langle n_{m,0,\sigma} \rangle \approx \frac{1}{6} )$. 
  Note that in doing so we have a model with, in total, $\frac{5}{3}$ electrons on the impurity.
  We only use density-density interactions and carry out the
  time evolution to $t=16$\,eV$^{-1}$. We set the truncated weight 
  to $t_w = 10^{-8}$, but restrict the bond dimension of the 
  impurity-impurity links to $\chi_{max}=200$.

\begin{figure}[ht]
   \centering
   %\scalebox{1}{ \input{tikzfigs/Gtau5Band} }
   \includegraphics{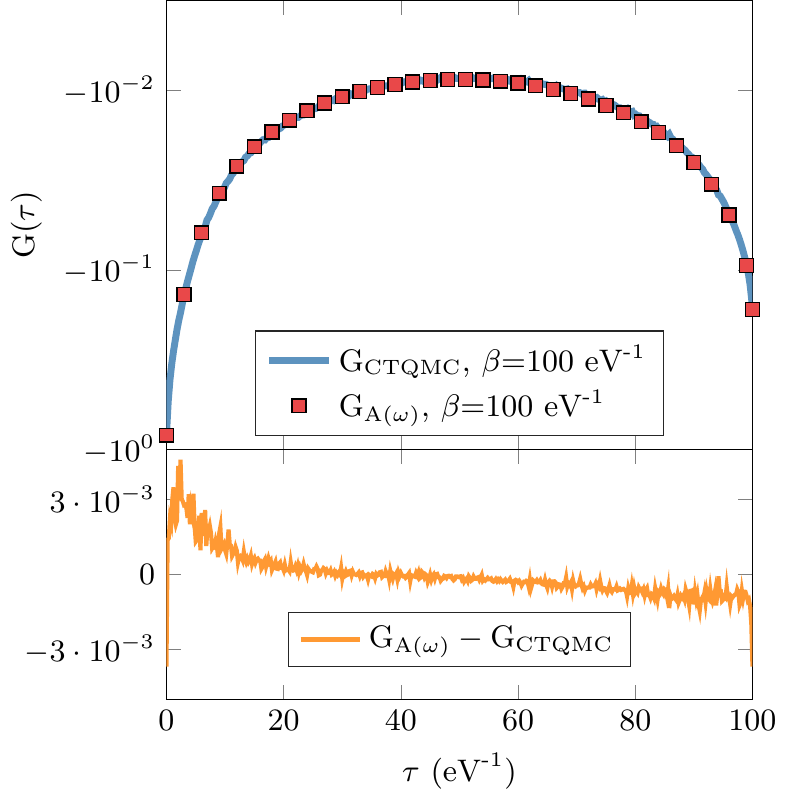}
   \caption{Comparison of imaginary time Greens functions $G(\tau)$
     from CTQMC (G$_{\text{CTQMC}  }$, blue line) and FTPS using Eq.~\eqref{eq:AOmToGtau} 
     ($\text{G}_{\text{A}(\omega) }$, red squares). As in Fig. \ref{fig:Gtau} they compare very well.} 
   \label{fig:Gtau5B}
 \end{figure}

  In Fig. \ref{fig:Gtau5B} we compare the results obtained for this five-band model
  to results from CTQMC, where we used the same discretized 
  bath hybridization as input to CTQMC. We again see excellent
  agreement, even on a logarithmic scale. The spectrum $A(\omega)$ (not shown) again exhibits strong
  structure in the upper Hubbard band. Of course, the computational 
  complexity is larger than in the three-orbital case and it grows during time evolution. Calculating the Greens 
  function took about 190 hours on the processors specified in Appendix~\ref{ssec:TrW}. We want to stress tough
  that the resulting spectrum (as well as Fig. \ref{fig:Gtau5B}) was already fully converged at $t=12$\,eV$^{-1}$
  (70 hours). We note that even with only one CPU hour ($t=6$\,eV$^{-1}$) the resulting spectrum is almost converged and barely 
  distinguishable from the final result. The benchmark therefore shows that with our
  FTPS approach a full five-orbital DMFT calculation is well within reach.

\section{Conclusions}
We have presented a novel multi-orbital impurity solver
which uses a fork-like tensor network whose
geometry resembles that of the Hamiltonian. The network structure is simple enough to generate Schmidt
decompositions, allowing us to truncate the 
tensor network safely and to use established methods like DMRG and real
time evolution. The solver works on the real frequency axis, and hence allows 
to formulate the full DMFT self
consistency procedure for real frequencies. Therefore, results are
not plagued by an ill-conditioned analytic continuation. Our approach exhibits no sign
problem, tough it does become more involved for larger numbers of orbitals. 

We tested the solver within DMFT on a Hamiltonian typically used for
the testbed material \ce{SrVO3} and investigated the influence of full
rotational invariance on the results.
We found clear spectral structures in particular in the upper
Hubbard band that have not been accessible by CTQMC, for which the necessary 
analytic continuation prohibits the resolution of fine structures in the spectral function at
higher energies. For our calculations with $U=4.0$\,eV, each peak in the
spectrum corresponds to an atomic excitation. Even excitations into
states with three particles on the impurity are resolved, as tiny spectral
peaks at high energies. Furthermore, upon increasing $U$, an additional
structure appears on the inside of the Hubbard bands, similar to
the precursors of the sharp Hubbard side peaks found for the one- and two-band Hubbard
models on the Bethe lattice~\cite{Ganahl2BHubb,GanahlCheby}. We have also shown 
that our approch is feasible for five-orbital models, by comparing results from 
the
FTPS solver to CTQMC for an artificial five-band model.

\section*{Acknowledgments}
The authors acknowledge financial support by the Austrian Science Fund
(FWF) through SFB ViCoM F41 (P04 and P03), through project P26220, and through 
the START program Y746, as well as by NAWI-Graz. 
This research was supported in part by the National Science Foundation under Grant No. NSF PHY-1125915.   
We are grateful for stimulating discussions with
F. Verstraete, K. Held, F. Maislinger and G. Kraberger. The
computational resources have been provided by the Vienna
Scientific Cluster (VSC). All calculations involving tensor networks
were performed using the ITensor library~\cite{ITensor}.

\appendix
\section*{\label{sec:Appendix}Appendix}

\begin{figure}[t!]
   \centering
   %\scalebox{1}{ \input{tikzfigs/LP_params} }
   \includegraphics{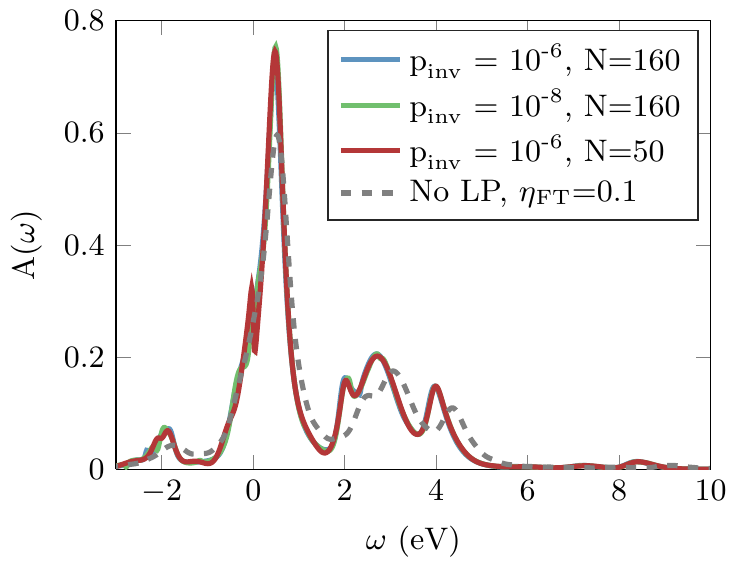}
   \caption{Spectrum $A(\omega)$ using different linear prediction (LP) parameters for a calculation without spin-flip and pair-hopping terms. The calculations with LP were performed with a broadening of $\eta_{FT}=0.02$\, eV. Except for small changes around $\omega=0$, the effect of the various LP parameters is minor. The blue line directly lies below the red and green line. We also show a DMFT calculation without any LP. Even then the main features are still present.
}
   \label{fig:LP}
  \end{figure}
\begin{figure}[t!]
   \centering
   %\scalebox{1}{ \input{tikzfigs/Cutoff} }
   \includegraphics{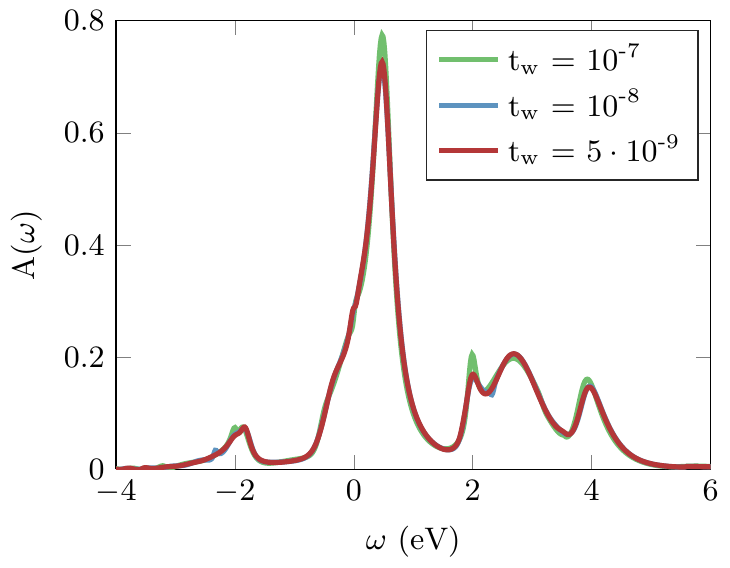}
   \caption{Different truncation values $t_w$ in the time evolution do not influence the shape of the spectrum $A(\omega)$.}
   \label{fig:Cutoff}
\end{figure}

In this appendix we show that our results
are very stable over a wide range of computational parameters. 
First we focus on the linear prediction method (Sec.~\ref{ssec:LP}). 
Then we show that the results are converged with
respect to the usual MPS-approximation~(Sec.~\ref{ssec:TrW}). 

\subsection{\label{ssec:LP}Linear Prediction}

 In order to obtain smooth and sharp spectra, we used linear prediction
 (LP) to extrapolate the Greens function in time~\cite{WhiteLinPred,Ganahl2BHubb}. 
 Without going into detail, we state the fact that linear prediction has 
 two parameters, the pseudo inverse cutoff $p_{\text{inv} }$ and the order $N$ of
 the linear prediction. Fig.~\ref{fig:LP} shows that 
the results are converged in these parameters.

We also show a DMFT run without any linear prediction, which is only 
possible if we increase the broadening parameter of the Fourier
transform to $\eta_{FT}=0.1$\,eV, since otherwise we would get
oscillations due to the hard cutoff of the time series. Except for a shift
towards the right, omitting the linear prediction only changes the
height (and width) of the peaks, but not the overall structure. This
is a strong indicator of the stability of these features.

\subsection{\label{ssec:TrW}Truncation of the Tensor Network}

One, if not the most important, parameter in any MPS-like calculation
is the sum of discarded singular values in each SVD (truncated weight
$t_w$). We want to emphasize that this parameter is the only
approximation in the representation of a state as a tensor-product
state, as we do not impose any hard cutoff on the bond
dimensions. Fig.~\ref{fig:Cutoff} shows that the spectrum is well
converged with respect to the truncation error during time
evolution. 

Finally, we want to comment on the required computational effort. In the
calculation without full rotational symmetry, the size of the largest
tensor to represent the ground state was\footnote{$35$ and $22$
correspond to the impurity links, $9$ is the bond dimension to the first bath 
site and $2$ is the physical bond dimension.}
$35\times22\times9\times2$ ($t_w=10^{-9}$) and at the end of time
evolution $127\times79\times30\times2$ ($t_w = 10^{-8}$). For a truncated weight of $t_w=10^{-7}$,
calculating the Greens function took about 17 minutes on a
node with two processors (Intel Xeon E5-2650v2, 2.6 GHz with 8 cores, and $G^>$ and $G^<$ each calculated on one processor). 
This time increases to five hours for the lowest truncated weight of $t_w=5\cdot10^{-9}$. 
Using the full rotationally invariant Hamiltonian, the biggest tensor in
the ground-state search was $90\times40\times10\times2$ ($t_w = 10^{-8}$) and
at the end of time evolution $79\times46\times21\times2$ ($t_w = 2\cdot10^{-8}$). 
The Greens function takes about three hours, and we need five hours for one 
full DMFT-cycle 
on the same two processors as above.

\bibliography{paper}

\end{document}